\documentclass[aps,prl,twocolumn,showpacs,groupedaddress,preprintnumbers,amsmath,amssymb]{revtex4}
\usepackage {graphicx}

\begin {document}

\title {Orbital Selective Pressure-Driven Metal-Insulator Transition in FeO from Dynamical Mean-Field Theory }

\author {A.~O.~Shorikov}
\affiliation {Institute of Metal Physics, Russian Academy of Sciences,
620990 Yekaterinburg, Russia\\
$^{2}$Ural Federal University, 
620002 Yekaterinburg, Russia}

\author {Z.~V.~Pchelkina}
\affiliation {Institute of Metal Physics, Russian Academy of Sciences,
620990 Yekaterinburg, Russia\\
$^{2}$Ural Federal University, 
620002 Yekaterinburg, Russia}

\author {V.~I.~Anisimov}
\affiliation {Institute of Metal Physics, Russian Academy of Sciences,
620990 Yekaterinburg, Russia\\
$^{2}$Ural Federal University, 
620002 Yekaterinburg, Russia}

\author {S.~L.~Skornyakov}
\affiliation {Institute of Metal Physics, Russian Academy of Sciences,
620990 Yekaterinburg, Russia\\
$^{2}$Ural Federal University, 
620002 Yekaterinburg, Russia}

\author {M.~A.~Korotin}
\affiliation {Institute of Metal Physics, Russian Academy of Sciences,
620041 Yekaterinburg, Russia}

\begin {abstract}

In this Letter we report the first LDA+DMFT (method combining Local 
Density Approximation with Dynamical Mean-Field Theory) results of magnetic and spectral properties calculation for paramagnetic phases 
of FeO at ambient and high pressures (HP). At ambient pressure (AP) calculation gave FeO as a Mott insulator with Fe 3$d$-shell in high-spin state. Calculated spectral functions are in a good agreement with experimental PES and IPES data.
Experimentally observed metal-insulator transition at high pressure is successfully reproduced in calculations. In contrast to MnO and Fe$_2$O$_3$ ($d^5$ configuration) where metal-insulator transition is accompanied by high-spin to low-spin transition, in FeO  ($d^6$ configuration) average value of magnetic moment
$\sqrt{<\mu_z^2>}$ 
 is nearly the same in the insulating phase at AP  and metallic phase at HP in agreement with  X-Ray spectroscopy data (Phys. Rev. Lett. {\bf83}, 4101 (1999)). The metal-insulator transition is orbital selective with only   $t_{2g}$ orbitals demonstrating  spectral function typical for strongly correlated metal (well pronounced Hubbard bands and narrow quasiparticle peak) while  $e_g$ states remain insulating.
\end {abstract}

\pacs {74.25.Jb, 71.45.Gm}

\maketitle
{\it Introduction.}--
For many years one of the central issues of condensed matter physics is the
metal-insulator transition (MIT) in $d$- or $f$-elements compounds~\cite{Imada}. The most spectacular examples are pressure-driven  transitions from wide gap Mott insulators to metallic state for transition metal oxides. For MnO and Fe$_2$O$_3$ ($d^5$ configuration)  metal-insulator transition is accompanied by high-spin to low-spin transition (HS--LS). Recently MIT in those materials was successfully described theoretically by LDA+DMFT(method combining Local 
Density Approximation with Dynamical Mean-Field Theory)~\cite{LDA+DMFT} calculations \cite{mno,fe2o3}.

Iron oxide also exhibits MIT under high pressure. Resistivity measurements  showed that FeO becomes metallic at pressures exceeding 72 GPa \cite{knittle}.
Correct description of MIT under pressure in w\"ustite (Fe$_{1-x}$O) is crucial in Earth science because iron oxides are believed to be major constituents of Earth mantle.  

At ambient pressure and room temperature FeO has cubic rocksalt B1 structure~\cite{Willis}. Below N\'eel temperature T$_N$=198~K FeO transforms into rhombohedral structure that could be viewed as a slight elongation along cube diagonal of the original cubic structure. Under pressure at room temperature rhombohedral distortion is observed at $\approx$ 15 GPa and this  structure is preserved up to at least 140 GPa \cite{yagi,ono}. This transformation to rhombohedral structure was believed to accompany long-range magnetic ordering due to increasing of N\'eel temperature with pressure \cite{okamoto}. However recent neutron diffraction study of w\"ustite at room temperature under pressure \cite{ding}  showed the absence of magnetic peaks corresponding to antiferromagnetism. 
At high pressures and temperatures ($P>$120 GPa and $T>$1000~K) FeO transforms into NiAs B8 phase~\cite{Fei}. 

In contrast to MnO and Fe$_2$O$_3$ it is not clear if  FeO  undergoes HS--LS transition with increase of pressure. Controversial experimental evidences were obtained for this problem.  M\"ossbauer spectroscopy~\cite{Pasternak} shows that quadrupole splitting appears between 60 and 90 GPa at room temperature. The authors interpreted that as LS diamagnetic state. On the other hand high pressure X-ray emission spectroscopy ~\cite{Badro} demonstrates that the satellite feature in Fe K$\beta$ line associated with HS Fe$^{2+}$  state does not disappear up to 143 GPa. 

Electronic structure calculations in standard Density Functional Theory (DFT) methods predict an antiferromagnetic metallic ground state \cite{isaak} in contrast to experimentally observed insulator with an optical band gap of 2.4 eV \cite{bowen}.The LDA+U method~\cite{Anisimov1991} has been successfully applied to investigate strongly correlated transition metal oxides and predicted an insulating ground state in FeO at ambient pressure \cite{feo-ldau}. Further investigation  done by Gramsch et al.~\cite{Savrasov} for stoichiometric w\"ustite has showed that using the value of  Coulomb parameter $U$ that reproduces experimentally observed energy gap at ambient pressure one can obtain  metal-insulator transition in LDA+U calculations for unrealistically high pressures only.

MIT in transition metal oxides with pressure can be successfully described using  LDA+DMFT  calculations \cite{mno,fe2o3}. In the present work we demonstrate that LDA+DMFT method reproduces MIT for FeO with pressure. However in contrast to  MnO and Fe$_2$O$_3$  MIT is not accompanied by high-spin to low-spin transition and metallic spectral function is observed only for $t_{2g}$ orbitals while $e_g$ states remain insulating.

{\it Method.}-- 
The LDA+DMFT method \cite{LDA+DMFT} calculation scheme is constructed in the following way: first, 
a Hamiltonian $\hat H_{LDA}$ is produced using converged LDA
results for the system under investigation, then the many-body
Hamiltonian is set up, and finally the corresponding self-consistent
DMFT equations are solved.
The calculations  presented below  have been done for crystal volumes corresponding to  values of pressure up to 140 GPa and room temperature. 
Since no structure transition has been observed at low temperatures~\cite{ono} and NiAs phase appears above 1000 K only all calculation were performed for simple NaCl (B1) cubic crystal structure with lattice constant scaled to give a volume corresponding to applied pressure \cite{knittle}. 
Ab-initio calculations of electronic structure  were   obtained within the pseudopotential plane-wave method PWSCF, as
implemented in the Quantum ESPRESSO package~\cite{PW}. Hamiltonians $\hat H_{LDA}$ in Wannier function (WF) basis~\cite{Wannier37, MarzariVanderbilt} were produced using projection procedure  that is  described in
details in Ref.~\onlinecite{Korotin}.
        
The WFs are defined by the choice of Bloch functions Hilbert space and by a
set of trial localized orbitals that will be projected on these Bloch
functions. The basis set includes all bands  that are formed by O-$2p$  and Fe-$3d$ states and
correspondingly full set of O-$2p$, and Fe-$3d$ atomic orbitals to
be projected on Bloch functions for these bands. That would correspond to
the extended model where in addition to $d$-orbitals all $p$-orbitals are
included too. 

The resulting 8x8 $p-d$ Hamiltonian to be solved by DMFT has the form
\begin{equation}
\hat H= \hat H_{LDA}- \hat H_{dc}+\frac{1}{2}\sum_{i,\alpha,\beta,\sigma,\sigma^{\prime}}
U^{\sigma\sigma^{\prime}}_{\alpha\beta}\hat n^{d}_{i\alpha\sigma}\hat n^{d}_{i\beta\sigma^{\prime}},
\label{eq:ham}
\end{equation}
where $U^{\sigma\sigma^{\prime}}_{\alpha\beta}$ is the Coulomb interaction matrix, 
$\hat n^d_{i\alpha\sigma}$ is the occupation number operator 
for the $d$ electrons with orbitals $\alpha$ or $\beta$ and spin indices $\sigma$ 
or $\sigma^{\prime}$ on the $i$-th site. 
The term $\hat H_{dc}$ stands for the {\it d}-{\it d} interaction 
already accounted for in LDA, so called double-counting correction. 
In the present calculation the double-counting was chosen in 
the following form $\hat H_{dc}=\bar{U}(n_{\rm dmft}-\frac{1}{2})\hat{I}$.
Here $n_{\rm dmft}$ is the self-consistent total number of {\it d} electrons 
obtained within the LDA+DMFT, $\bar{U}$ is the average Coulomb parameter for 
the {\it d} shell and $\hat I$ is unit operator. 

The elements of $U_{\alpha\beta}^{\sigma\sigma'}$ matrix
are parameterized by $U$ and $J_H$ according to procedure described in
\cite{LichtAnisZaanen}. The values of Coulomb repulsion parameter $U$ and Hund exchange parameter $J_H$ were calculated by the constrained LDA method \cite{U-calc} on Wannier functions \cite{Korotin}. Obtained values  $J_H$=0.89 eV,  $U$ = 5 eV are close to  previous estimations~\cite{Savrasov}.   
The effective  impurity problem for the DMFT was solved
by the hybridization  expansion Continuous-Time Quantum Monte-Carlo 
method (CT-QMC) \cite{CTQMC}.  Calculations for all volumes were performed 
in the paramagnetic state at the inverse temperature $\beta=1/T$ = 40 eV$^{-1}$ corresponding to 290~K. Spectral functions on real energies  were calculated by Maximum Entropy Method (MEM)\cite{mem}.

{\it Results and discussion.}--
The Fe $d$ band is split by crystal field in triply degenerated $t_{2g}$ and doubly degenerated $e_g$ subbands. LDA fails to describe insulating ground state of FeO at AP and for all volumes FeO is metallic. 
\begin {figure}
\vspace{5mm}
\includegraphics [width=0.425\textwidth]{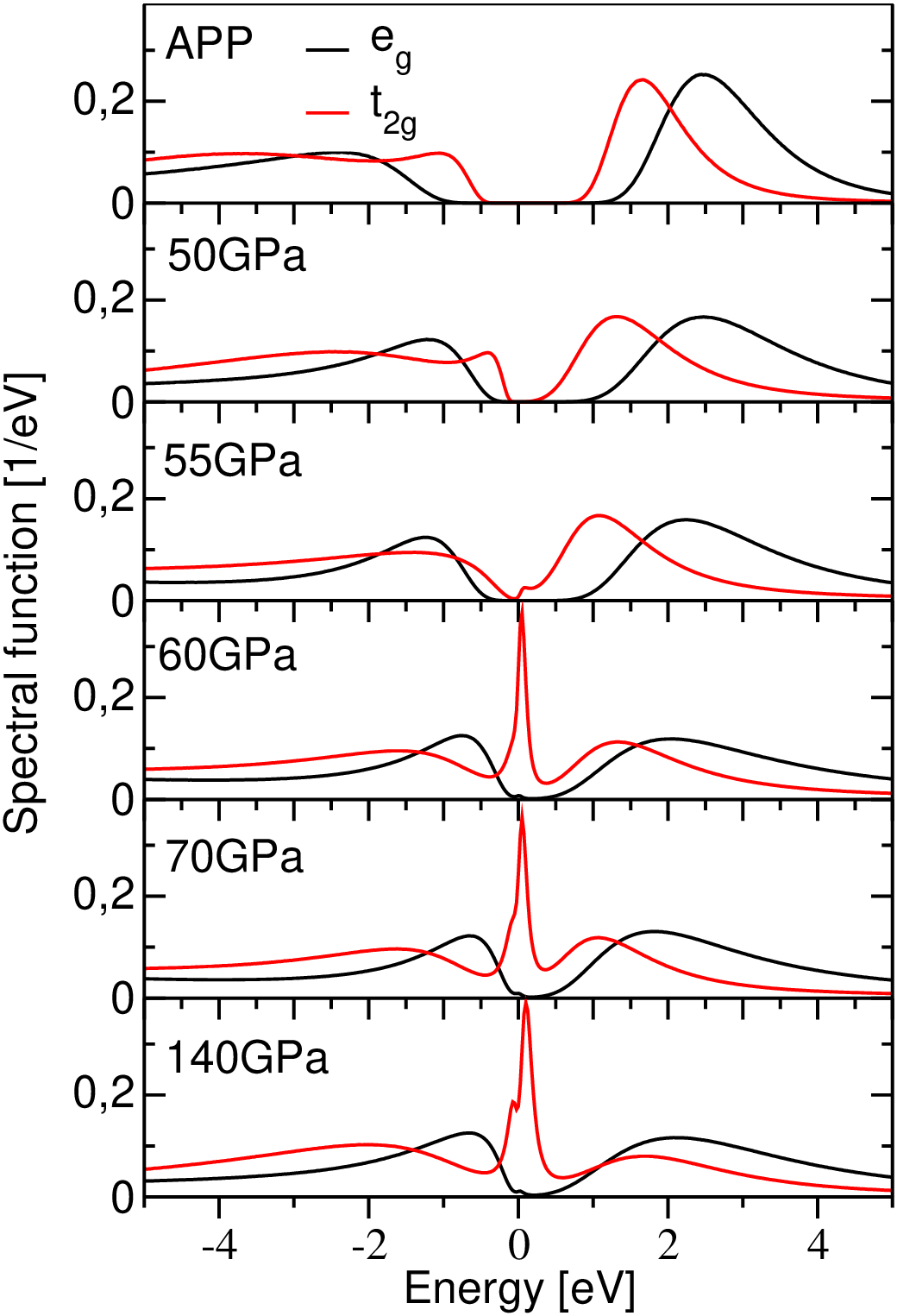}
\caption {(Color online) Spectral function of Fe-d states vs. pressure obtained in LDA+DMFT (CT-QMC) calculations at room temperature.}
\label {fig:ins_sf}
\end {figure}

\begin {figure}
\vspace{5mm}
\includegraphics [width=0.425\textwidth]{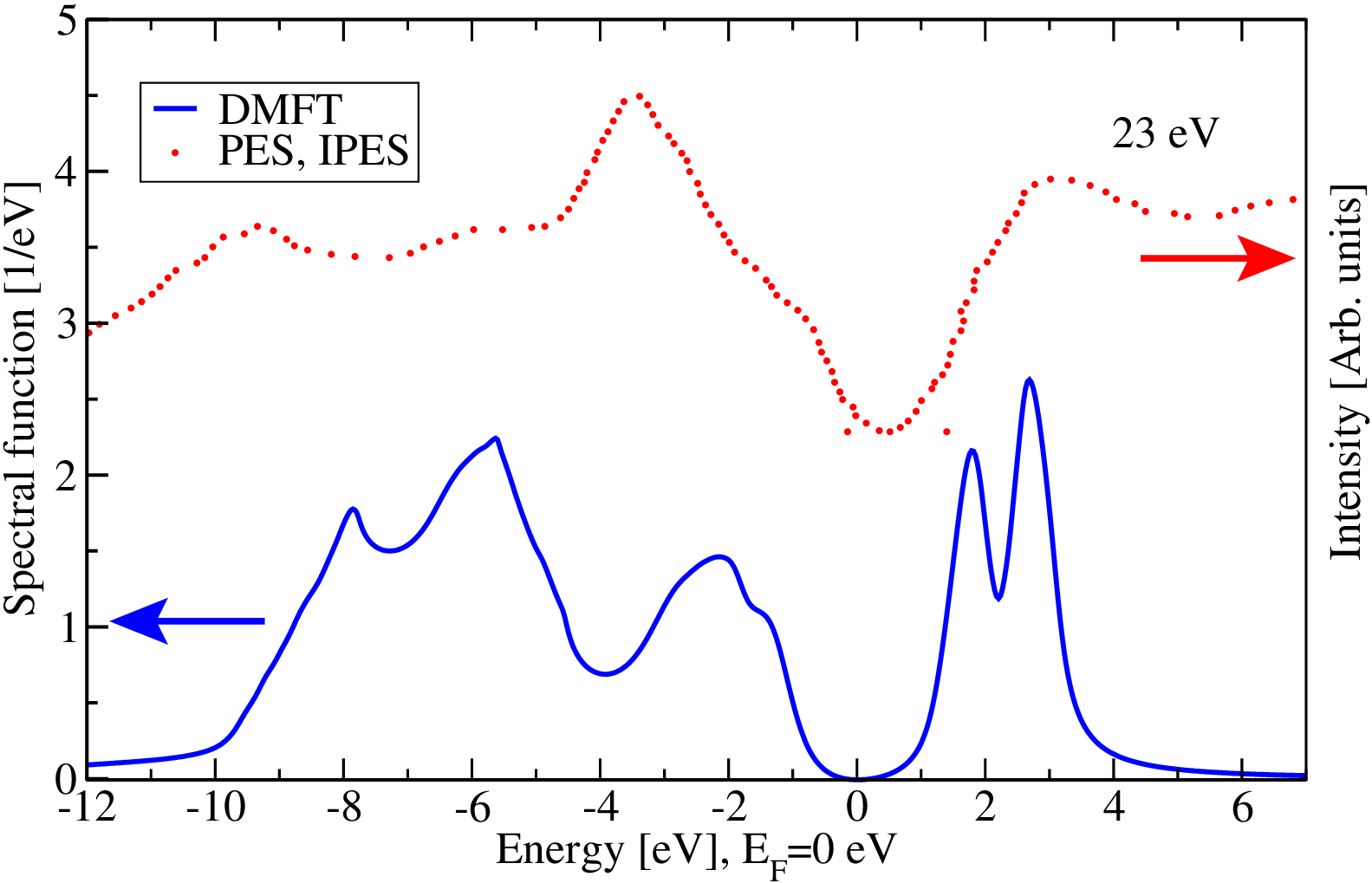}
\caption {(Color online) Total spectral function of FeO in ambient pressure phase calculated within LDA+DMFT (CT-QMC) ($\beta$=40 eV$^{-1}$)(solid blue line) in comparison with combined PES and IPES experimental data (red dots) from Ref.~\cite{Fujimori, Braicovich}.}
\label {fig:app_exp}
\end {figure}

Including Coulomb correlation effects in frames of LDA+DMFT  method results in    high spin state wide gap Mott insulator  for ambient pressure phase (APP) of FeO  in agreement with experimental data. The calculated energy gap value of about 2 eV agrees well with IPES measurement~\cite{Kim} value 2.5 eV and optical spectrum~\cite{bowen} value 2.4 eV.  The  occupation numbers  for Fe d orbitals are n($e_{g}$)=0.54 and n($t_{2g}$)=0.68. The average value of  local magnetic moment $\sqrt{<\mu_z^2>}$ is 3.8$\mu_B$. Those numbers agree very well with high-spin state of Fe$^{+2}$ ion (d$^6$ configuration) in cubic crystal field: 2 electrons in $e_{g}$ states (n($e_{g})$=1/2) and 4 electrons in $t_{2g}$ states (n($t_{2g}$)=2/3) with magnetic moment value 4 $\mu_B$.
Spectral functions $A(\omega)$ for all pressure values calculated by MEM 
using Green function $G(\tau)$  from CT-QMC calculations are presented in the Fig.~\ref{fig:ins_sf}. The spectral function for ambient pressure phase (APP) shows well defined insulating behavior for all $d$-orbitals. However the energy gap for $e_{g}$ states is nearly two times larger than for $t_{2g}$ states indicating that the latter orbitals are closer to MIT than the former ones.
Figure~\ref{fig:app_exp} contains calculated total spectral function compared with spectrum combined from PES and IPES experiments~\cite{Fujimori, Braicovich}. The theoretical and experimental curves are in a good agreement.  

\begin {figure}
\vspace{5mm}
\includegraphics [width=0.425\textwidth]{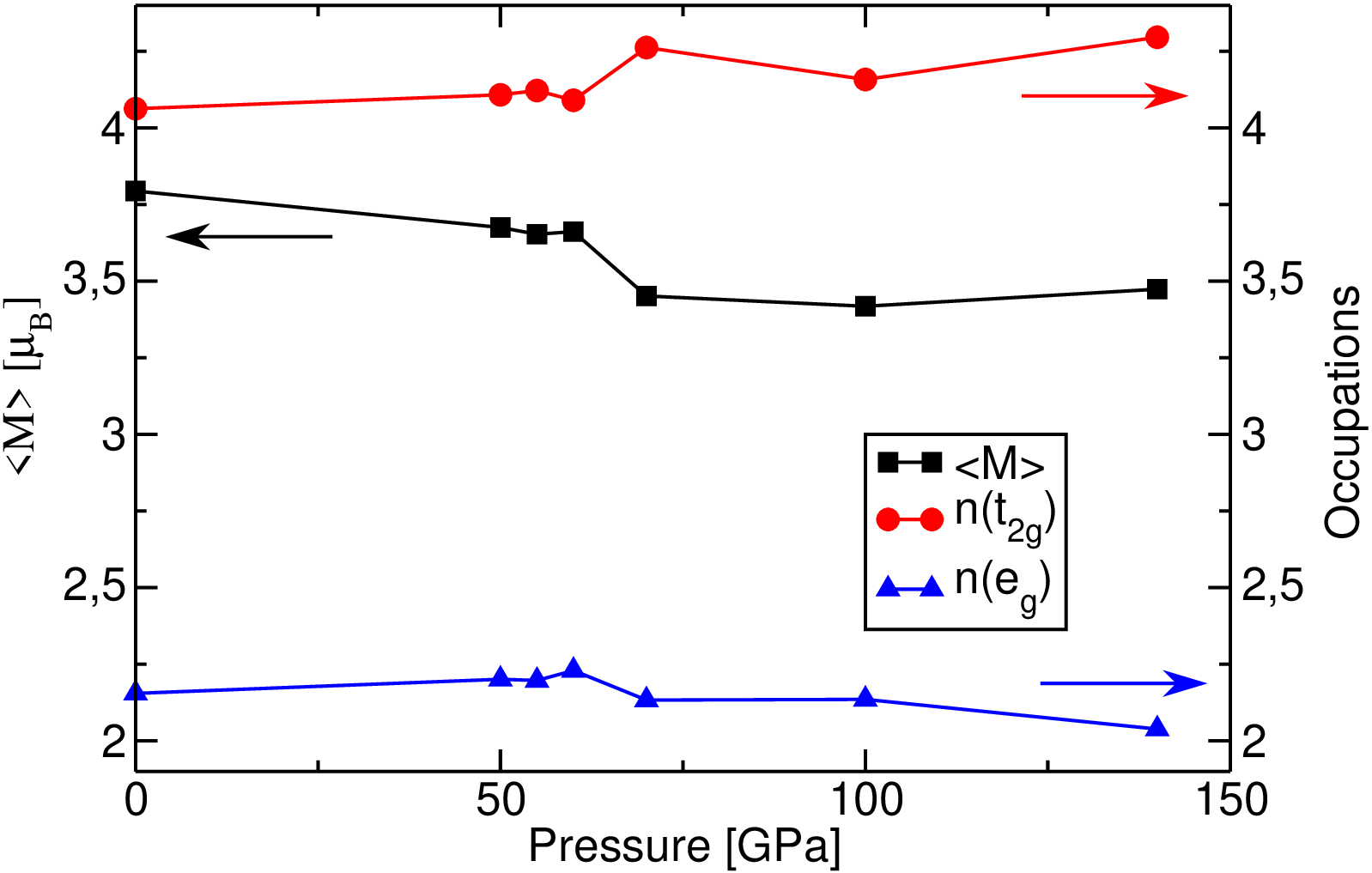}
\caption {(Color online) Magnetic moments (black squares) and occupancies of $t_{2g}$ (red cicles) and $e_g$ (blue triangles) shells vs. pressure obtained in LDA+DMFT (CT-QMC) calculations. }
\label {fig:met}
\end {figure}

LDA+DMFT calculation made for small volume values corresponding to high pressures  gave  metallic state for  FeO (see Fig.~\ref {fig:met}) starting from 60 GPa in  agreement with experiment ~\cite{knittle}.
One can see that $t_{2g}$ orbitals become metallic whereas $e_g$ ones remain insulating. This behavior reminds the orbital selective Mott transition (OSMT) in ruthenates \cite{osm}. Occupation number values in Fe-$d$ shell are practically not changed comparing with APP and are n($e_{g}$)=0.55 n($t_{2g}$)=0.68 at 140~GPa. The magnetic moment value decreases on a few percent only and is 3.5$\mu_B$ at 140~GPa. The only interpretation for those values is that  an iron $d$-shell in high pressure metallic phase of FeO  still corresponds to high-spin state of Fe$^{+2}$ ion.
This conclusion agrees well with analysis of   high pressure X-Ray emission spectroscopy experiment made in Ref.~\cite{Badro}. The occupation numbers and magnetic moment vs. pressure are presented in the Fig.~\ref{fig:met}. One can see that all curves exhibit the kink at 60~GPa. We argue that this feature is due to MIT and corresponding reconstruction of spectral function at Fermi level. 
Spectral functions $A(\omega)$ for $t_{2g}$ in the Fig.~\ref {fig:ins_sf} for pressure values larger then 60 GPa become typical for strongly correlated metal close to MIT: well pronounced Hubbard bands and narrow quasiparticle peak. $A(\omega)$ for $e_{g}$ is still insulating with Hubbard bands only but energy gap value is strongly decreased comparing with APP (see Fig.~\ref{fig:ins_sf}).

\begin {figure}
\vspace{5mm}
\includegraphics [width=0.425\textwidth]{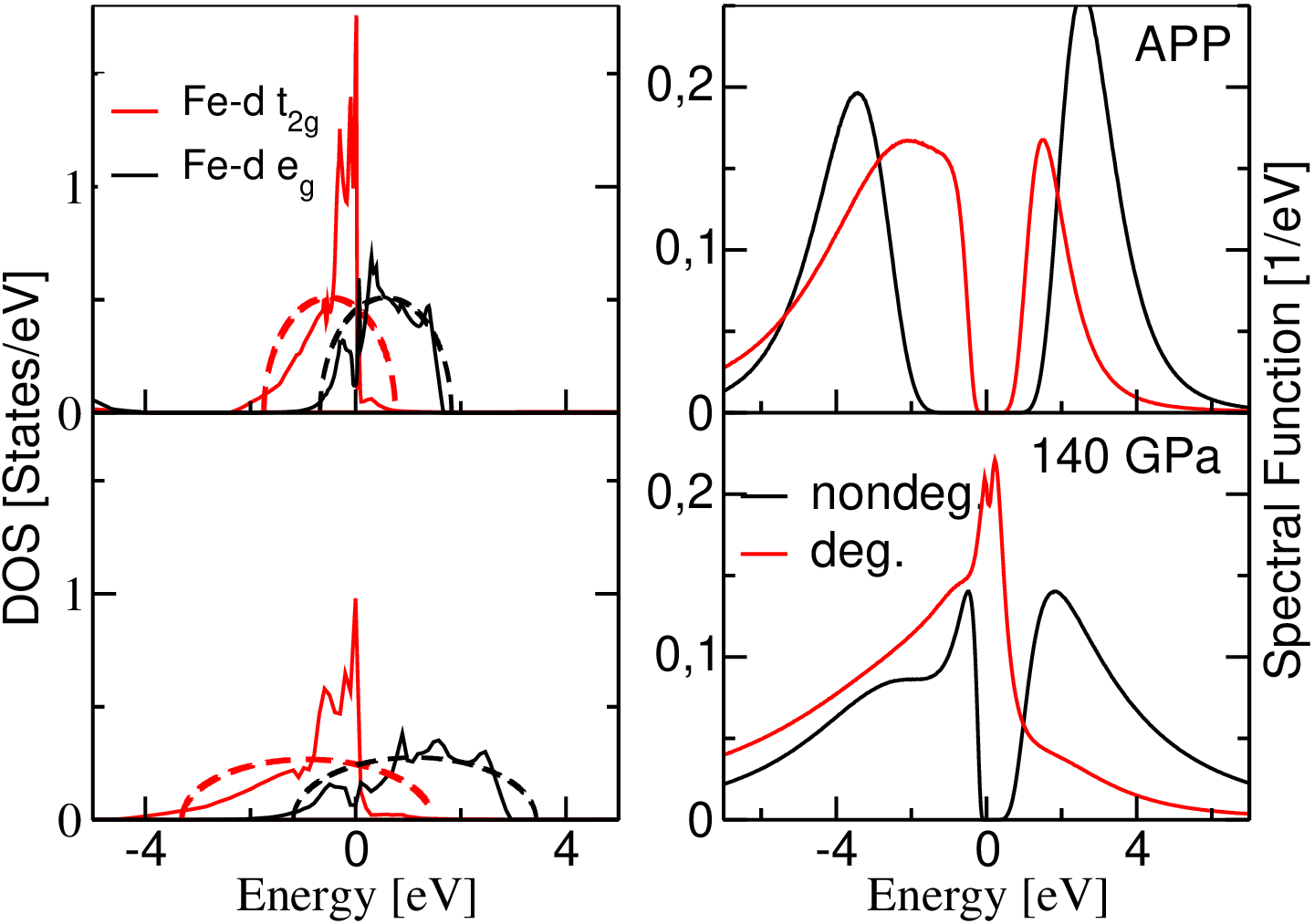}
\caption {(Color online) Left panels - LDA densities of states (DOS) (solid lines) and correponding model semicircle DOS(dashed lines). Right panels - spectral functions from model DMFT (CT-QMC) calculations for two values of pressure. Non degenerate orbital (black lines) reproduces $e_g$-orbitals and  2 times degenerate one reproduces  $t_{2g}$ orbitals (red line).}
\label {fig:model}
\end {figure}

To understand these results the following simple model was used. The model has two semicircle DOS of the same width with 3 orbital  and 4 electrons. One orbital is non-degenerate and two other orbitals are degenerate. The centers of gravity and DOS  widths were taken from \emph{ab initio} LDA calculations. In this model non-degenerate orbital stands for $e_g$-orbital in FeO and two others for $t_{2g}$. Occupations in model in HS state  are 1/2 for non-degenerate orbital (the same as in realistic LDA+DMFT calculation for FeO) and 3/4  for degenerate orbitals comparing with 2/3 in the case of $t_{2g}$ orbitals in FeO. The Kanamori parametrisation of Coulomb repulsion (with the same U=5 eV and J=0.89 eV) was used. Note, that corresponding matrix elemetns $U^{\sigma,\sigma'}_{\alpha,\beta}$ (eq.~\ref{eq:ham}) are set to be the same for all orbitals. The model was solved using DMFT (CT-QMC) method and obtained spectral functions for two values of pressure (APP and 140 GPA) are presented in the Fig.~\ref{fig:model}. 

The orbital selective metal-insulated transition (OSMT) was reproduced in these calculations. Since DOSes for all three orbitals have  the same width (in contrast to OSMT \cite{osm} in ruthenates where two bands have very different widths) and actual structure of DOS is neglected we can conclude that effects of different degeneracy of orbitals and deviation from half filling  are the driving force of this separate transition. It is known that critical value of Coulomb interaction parameter $U_c$ needed for metal-insulator transition in half-filled degenerate Hubbard model is $U_c\approx \sqrt{N}U_c^{N=1}-N J$ \cite{han} ($N$ is degeneracy and $U_c^{N=1}$ is critical $U$ value for non-degenerate case). That means that for more degenerate t$_{2g}$ orbitals one needs larger effective $U$ value to become insulating than for less degenerate $e_g$ orbitals. In addition to that for half-filled states an estimation for effective $U_{eff}$ value is $U+(N-1)J$ while for the occupancy one electron more then half-filling an estimation is $U_{eff}=U-J$. Then for 2/3 filled $t_{2g}$ orbitals one needs much larger $U$ value to drive them into insulating state than for half-filled $e_g$ states.

\emph{In conclusion}.--
 We have performed LDA+DMFT calculation for FeO at room temperature  and  values of pressure from the ambient one till 140 GPa.  In the agreement with experiment spectral function for FeO at AP demonstrates an energy gap of about 2 eV. At the pressures higher then 60 GPa FeO is metallic but only for t$_{2g}$ orbitals while e$_g$ states remain insulating that corresponds to orbital selective Mott transition scenario. The MIT obtained in our calculations is not accompanied by change of spin state and FeO has HS with large local moment in  APP and all HPP. This result agrees with high pressure X-Ray emission spectroscopy data.       

{\it Acknowledgments.}-- 
The authors thank J.~Kune\v{s} for providing DMFT computer
code used in our calculations, P.~Werner for the CT-QMC impurity solver. 
This work was supported by the Russian Foundation for
Basic Research (Project No. 10-02-00046-a, 09-02-00431-a, and 10-02-00546-a), 
the Dynasty Foundation, the fund of the President of the Russian Federation 
for the support of scientific schools NSH 1941.2008.2, the Programs of
the Russian Academy of Science Presidium ``Quantum microphysics of
condensed matter'' N7 and ''Strongly compressed materials``, Russian Federal Agency for Science and Innovations 
(Program ``Scientific and Scientific-Pedagogical Trained of the Innovating 
Russia'' for 2009-2010 years), grant No. 02.740.11.0217, MK-3758.2010.2.


\begin {thebibliography}{99}
\bibitem {Imada} M. Imada, A Fujimori and Y. Tokura, Rev. Mod. Phys. 
\textbf {70}, 1039 (1998).
\bibitem {LDA+DMFT} V. I. Anisimov
 {\it et al.}, J. Phys.: Condens. Matter \textbf {9}, 7359
(1997); A. I. Lichtenstein and M. I. Katsnelson, Phys. Rev. B \textbf {57},
6884 (1998); K. Held
{\it et al.}, Phys. Stat. Sol. (b) \textbf {243}, 2599 (2006).
\bibitem{mno}	Jan Kunes {\it et al},
Nature Materials \textbf {7}, 198 (2008).

\bibitem{fe2o3}	J. Kunes {\it et al}, 
Phys. Rev. Lett. \textbf {102}, 146402 (2009).

\bibitem{knittle} E.Knittle, R. Jeanloz, A.C. Mitchel, and W.J. Nellis, Sol. State Comm. \textbf {59} 513 (1986).


\bibitem {Willis}B.T.M Willis and H.P.Rooksby, Acta Crystallographica \textbf {6}, 827-831 (1953). 

\bibitem{yagi} T. Yagi, T. Suzuki and S. Akimoto, J. Geophysical Research B  \textbf {90}, 8784 (1985).
\bibitem{ono}S. Ono, Y. Ohishi, T. Kikegawa, J. Phys.: Condens. Matter \textbf {19}, 3, 036205 (2007). 
\bibitem{okamoto} T. Okamoto, T. Fuji, Y. Hidaka and E. Hiroshima, J. Phys. Soc. Jpn, \textbf{23}, 1174 (1967).

\bibitem{ding} Y. Ding {\it et al},
Applied Phys. Lett. \textbf{86}, 052505 (2000).

\bibitem {Fei}Y. Fei and H.K. Mao, Science \textbf {226}, 1678-1680 (1994). 
\bibitem {Pasternak} M.P. Pasternak {\it et al}, Phys. Rev. Lett\textbf{79},
 5046 (1997). 
 \bibitem {Badro} Badro J. {\it et al}, Phys. Rev. Lett \textbf{83},
 1401 (1999). 
 
\bibitem{isaak}  D. Isaak, R. Cohen, M. Mehl and D. Singh, Phys.Rev. B \textbf {47},7720 (1993).

\bibitem{bowen} H. Bowen, D. Adler, and B. Auker, J. Sol. State Chem. \textbf{12}, 355 (1975).
 
\bibitem {Anisimov1991} V.I. Anisimov, J. Zaanen and O.K. Anderson, Phys. Rev. B \textbf {44}, 943 (1991).

\bibitem{feo-ldau} I. Mazin, V. Anisimov, Phys.Rev.B \textbf{55},  12 822 (1997).

\bibitem {Savrasov} S.A. Gramsch, R.E. Cohen and S.Yu. Savrasov, Amer. Miner. \textbf {88}, 257-261 (2003). 


\bibitem {PW} S. Baroni, S. de Gironcoli, A. D. Corso, and P. Giannozzi,
http://www.pwscf.org.


\bibitem {Wannier37} G. H. Wannier, Phys. Rev. \textbf {52}, 191 (1937).

\bibitem {MarzariVanderbilt} N. Marzari and D. Vanderbilt, Phys. Rev. B
\textbf {56}, 12847 (1997); W. Ku, H. Rosner, W. E. Pickett, and R. T.
Scalettar, Phys. Rev. Lett. \textbf {89}, 167204 (2002).

\bibitem{Korotin} Dm. Korotin {\it et al}.,
Euro. Phys. J. B {\bf 65}, 1434 (2008).

\bibitem{LichtAnisZaanen} A. I. Liechtenstein, V. I. Anisimov, and J. Zaanen,
Phys. Rev. B {\bf 52}, R5467 (1995).

\bibitem {U-calc} P. H. Dederichs, S. Bl\"ugel, R. Zeller, and H. Akai,
Phys. Rev. Lett. \textbf {53}, 2512 (1984); O. Gunnarsson, O. K. Andersen,
O. Jepsen, and J. Zaanen, Phys. Rev. B \textbf {39}, 1708 (1989); V. I.
Anisimov and O. Gunnarsson, \textit {ibid.} \textbf {43}, 7570 (1991).



\bibitem {CTQMC} P. Werner {\it et al}., 
Phys. Rev. Lett. {\bf 97}, 076405 (2006).

\bibitem{mem} Mark Jarrell
and J. E. Gubernatis, Phys. Rep. {\bf 269}, 133 (1996).

\bibitem {Kim} Bong-soo Kim {\it et al},  Phys. Rev. B \textbf{41},
 12227 (1989). 
 

\bibitem {Fujimori} A. Fujimori{\it et al}, Phys. Rev. B \textbf{36},
 6691 (1987).  
\bibitem {Braicovich} L. Braicovich {\it et al}, Phys. Rev. B \textbf{46},
 12165 (1992).  
\bibitem {osm} V.~I.~Anisimov \emph {et
al.}, Eur. Phys. J. B \textbf {25}, 191 (2002).

\bibitem{han} J. Han, M. Jarrell, D. Cox, Phys. Rev. B \textbf{58}, 4199 (1998).
 




\end {thebibliography}
\end {document}